\documentclass[%
 reprint,
superscriptaddress,
 amsmath,amssymb,
pre,
]{revtex4-2}

\usepackage{graphicx}% Include figure files
\usepackage{dcolumn}% Align table columns on decimal point
\usepackage{bm}% bold math
\usepackage{color}
\usepackage[colorlinks=true,allcolors=blue,breaklinks=true]{hyperref}
\usepackage{physics}
\usepackage{siunitx}

\begin{document}

\preprint{APS/123-QED}

\title{Tuning the Size and Stiffness of Inflatable Particles}% Force line breaks with \\

\author{Nidhi Pashine}
\thanks{These authors contributed equally to this work.}%
\affiliation{Department of Mechanical Engineering, Yale University, New Haven, Connecticut 06520, USA.}
\affiliation{Department of Physics and BioInspired Institute, Syracuse University, Syracuse, New York 13244, USA.}%Lines break automatically or can be forced with \\

\author{Dong Wang}%
\thanks{These authors contributed equally to this work.}%
\affiliation{Department of Mechanical Engineering, Yale University, New Haven, Connecticut 06520, USA.}

\author{Robert Baines}
\affiliation{Department of Mechanical Engineering, Yale University, New Haven, Connecticut 06520, USA.}

\author{Medha Goyal}
\affiliation{Department of Mechanical Engineering, Yale University, New Haven, Connecticut 06520, USA.}

\author{Mark D. Shattuck}
\affiliation{Department of Physics and Benjamin Levich Institute, City College of New York, New York City, New York 10031, USA.}

\author{Corey S. O'Hern}
\affiliation{Department of Mechanical Engineering, Yale University, New Haven, Connecticut 06520, USA.}
\affiliation{Department of Physics, Yale University, New Haven, Connecticut 06520, USA. }
\affiliation{Department of Applied Physics, Yale University, New Haven, Connecticut 06520, USA. }
\affiliation{Graduate Program in Computational Biology \& Biomedical Informatics, Yale University, New Haven, Connecticut 06520, USA. }

\author{Rebecca Kramer-Bottiglio}
\affiliation{Department of Mechanical Engineering, Yale University, New Haven, Connecticut 06520, USA.}
\email{rebecca.kramer@yale.edu}

\date{\today}% It is always \today, today,
             %  but any date may be explicitly specified

\begin{abstract}
We describe size-varying cylindrical particles made from silicone elastomers that can serve as building blocks for robotic granular materials. The particle size variation, which is achieved by inflation, gives rise to changes in stiffness under compression. We design and fabricate inflatable particles that can become stiffer {\it or} softer during inflation, depending on key parameters of the particle geometry, such as the ratio of the fillet radius to the wall thickness, $r/t$. We also conduct numerical simulations of the inflatable particles and show that they only soften during inflation when localization of large strains occurs in the regime $r/t \rightarrow 0$. This work introduces novel particle systems with tunable size and stiffness that can be implemented in numerous soft robotic applications.

\end{abstract}

%\keywords{Suggested keywords}%Use showkeys class option if keyword
                              %display desired
\maketitle

\section{Introduction}

Granular materials composed of discrete frictional particles possess a host of complex yet tunable mechanical properties that have been the subject of study for decades~\cite{behringer_jamming_2015,liu_jamming_2010, behringer_physics_2018}. Granular metamaterials can be designed to exhibit targeted mechanical responses, such as phononic band gaps and the ability to propagate solitons and edge modes~\cite{zhang_designing_2023,gaspar_granular_2010, miskin_adapting_2013}. 
%Although granular metamaterials are typically composed of passive particles~\cite{pashine_tessellated_2023,fu_programmable_2019, haver_elasticity_2024, kocharyan_development_2023}, recent work has realized granular metamaterials consisting of particles that themselves have tunable properties, unlocking further functionality~\cite{parsa_evolution_2022, wu_active_2019}. 
Although granular metamaterials are typically composed of homogeneous particles with identical properties~\cite{pashine_tessellated_2023,fu_programmable_2019, haver_elasticity_2024, kocharyan_development_2023}, recent work has realized granular metamaterials where properties of individual particles can be varied, unlocking further functionality~\cite{parsa_evolution_2022, wu_active_2019}. 
To maximize the tunability of granular metamaterials, we need a diverse range of particles capable of dynamically varying properties such as size, shape, and stiffness within a granular packing.

%However, these particle properties are often pre-determined before being assembled into the granular metamaterials. There exist few studies in which particle properties can be tuned once forming the metamaterials.

In this work, we design and fabricate inflatable particles with tunable size and stiffness, which can form the basis for future studies of robotic granular materials. 
Tuning the individual particle properties in a robotic granular packing could provide control over the interparticle contact network, which will enable the programming of the global mechanical properties of the packings. 
Previous studies have incorporated changes in particle stiffness and actuation~\cite{li2025vs2}.
%~\cite{althoefer_antagonistic_2018} (e.g. using jamming in granular materials and phase-changing alloys~\cite{yang_reprogrammable_2021,tonazzini_variable_2016, santoso_single_2019, manti_stiffening_2016}). 
A unique contribution of the current work is the ability to control the stiffness-inflation pressure relationship of an inflatable particle. Specifically, we can control whether the particle's compressive stiffness increases or decreases with volumetric strain by adjusting key geometric parameters, which allows the particles to either stiffen or soften during inflation.

\begin{figure*}
    \centering
    \includegraphics[width=0.7\linewidth]{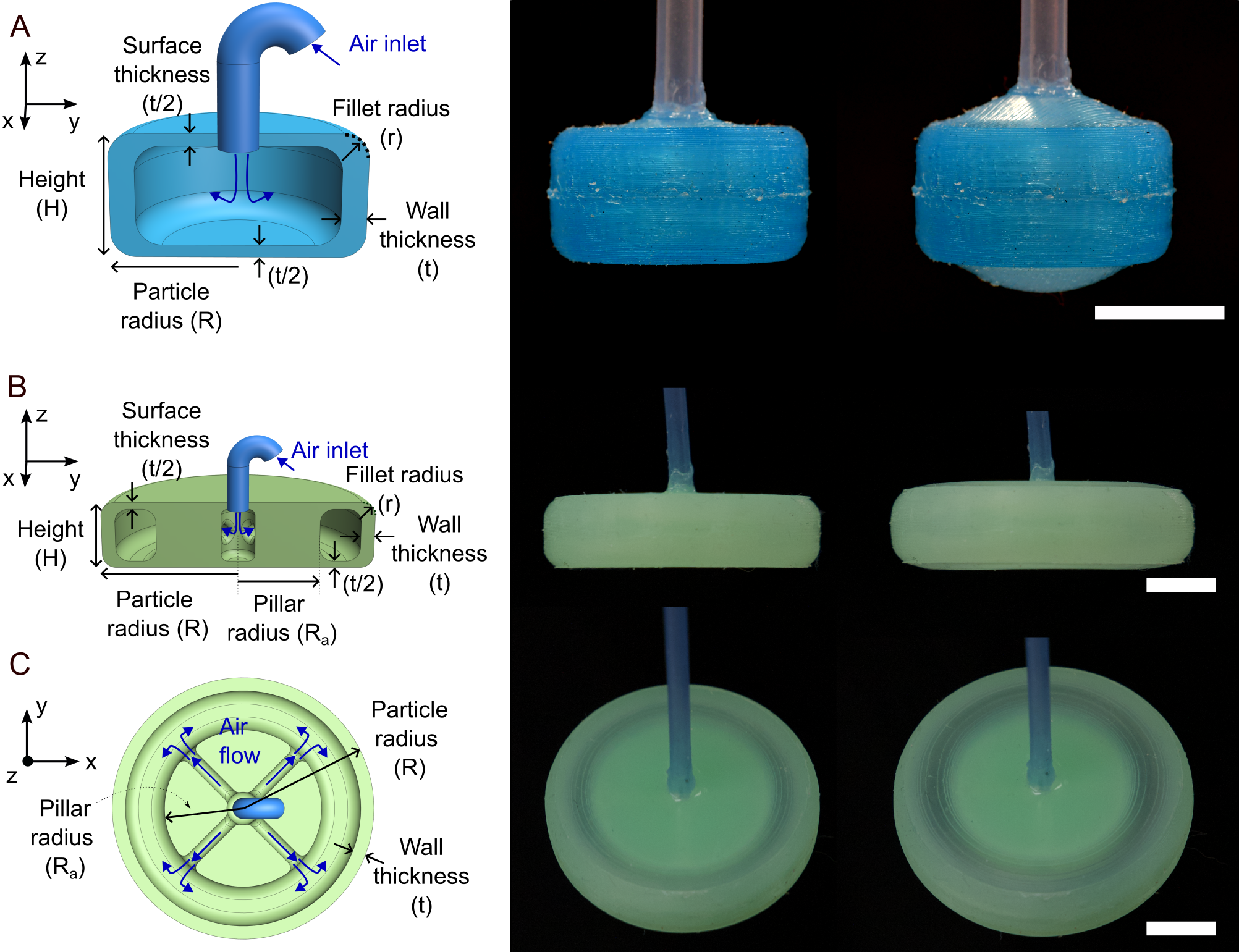}
    \caption{Schematic of the two types of particles (left) with images in the uninflated (middle) and inflated (right) states. Side views of (A) type I and (B) type II particles. (C) Top view of a type II particle. The type I and II particles have been inflated to $\sim 3.4 \unit{\kilo \pascal}$  and $\sim 8.6 \unit{\kilo \pascal}$, respectively. All scale bars correspond to 10~mm.}
    \label{fig:schematic}
\end{figure*}

This study focuses on two types of inflatable cylindrical particles: type I: cylindrical shells and type II: toroidal shells. Type I particles are empty inside and expand radially, as well as along the cylindrical axis of the particle, when inflated. In contrast, type II particles mostly expand radially, with minimal deformation along the cylindrical axis. Images of the two particle types are shown in Fig.~\ref{fig:schematic}. We limit the expansion of type II particles along the cylindrical axis by connecting the top and bottom surfaces with a central pillar.

The particle geometries are specified by several dimensionless ratios. In this work, we vary the shapes of both types of particles mainly by changing the ratio of the cylinder's corner radius ($r$ in Fig.~\ref{fig:schematic}) to its thickness ($t$ in Fig.~\ref{fig:schematic}). We measure the force response as a function of pressure when these particles are compressed radially. Studying these two particle designs allows us to determine how the particle deformation during inflation affects its compressive stiffness. 

For both types of particles, we find that the compressive stiffness can increase, decrease, or stay the same during inflation. With finite element method simulations, we show that the slope of the stiffness versus inflation pressure curve is negative only when substantial strain localization occurs near the cylinder's ``corners'' (where the top wall meets the side walls) during inflation, resulting in strain softening. 
%Finally, we show that the specific mode of particle deformation---either cylindrical or radial---during inflation does not strongly affect the stiffness variation since type I and type II particles possess similar stiffness versus inflation pressure behavior. 
We also demonstrate that the mode of particle deformation—whether cylindrical or radial—during inflation has little impact on stiffness variation, as both type I and type II particles exhibit similar stiffness behavior across different inflation pressures.
These findings are used to design particles with specific variation of the compressive stiffness with inflation pressure, \textit{i.e.}, increasing, decreasing, or constant stiffness. 

The remainder of the article is organized as follows. In Sec.~\ref{sec:methods}, we describe the manufacturing processes for type I and type II particles.  We also describe the finite element method (FEM) simulations conducted to understand the compressive stiffness for type I particles during inflation. In Sec.~\ref{sec:results}, we discuss experimental measurements of the particle stiffness versus the inflation pressure for type II particles. In Sec.~\ref{sec:conclusions}, we summarize the conclusions and propose promising directions for future research. In Appendix A, we show the experimental results for the relation between the change in radius of type II particles and the inflation pressure.

\section{Methods}
\label{sec:methods}

\subsection{Particle Fabrication}

\begin{figure}[ht!]
    \centering
    \includegraphics[width=\linewidth]{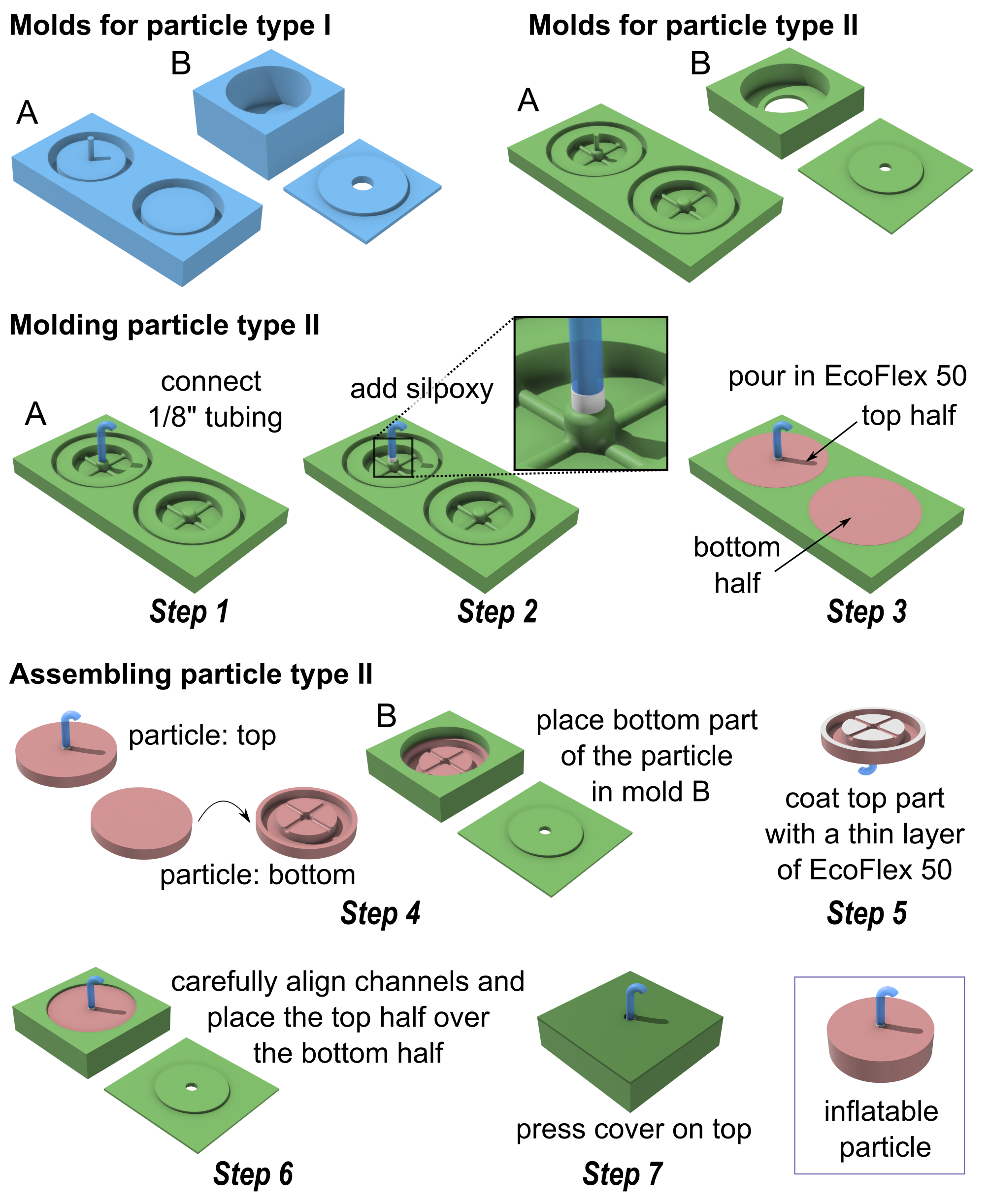}
    \caption{Summary of the particle fabrication process. Each type of particle requires two sets of molds: one for molding the two halves (mold A) and one for assembling the two halves together (mold B). The components of the molds are 3D-printed out of Polyvinyl Lactic Acid (PLA). The blue (green) molds are used for type I (type II) particles. The fabrication process for both particle types is nearly identical, except for the use of slightly different mold designs. The seven-step fabrication process for type II particles is shown. The particles (pink) are molded from EcoFlex\texttrademark~50.}
    \label{fig:fabrication}
\end{figure}

The inflatable particles are fabricated from EcoFlex\texttrademark~50, a two-part elastomer that reaches a 00-50 shore hardness upon curing. A schematic of the fabrication process is shown in Fig.~\ref{fig:fabrication}. The particles are molded as two halves and then attached to each other. The molds for each of the parts and for the final assembly are 3D-printed in PLA on a Prusa\texttrademark~MK3S+ printer.  The fabrication process involves first molding the top and bottom halves (mold A in Fig.~\ref{fig:fabrication}) and then assembling the two parts using mold B in Fig.~\ref{fig:fabrication} to make the particle. During the molding process, a silicone tube is attached to the top part of the particle using Silpoxy\texttrademark, which serves as an air inlet. After the two parts of the particle are cured, they are glued using a thin layer of uncured EcoFlex\texttrademark~50 and pressed in place using the second part of the mold. This particle assembly process ensures sufficient contact between the two particle layers so that leaks are avoided. 

The geometry of the uninflated type I particles is characterized by four dimensions: the height $H$, particle radius $R$, wall thickness $t$, and fillet radius $r$, as shown in Fig.~\ref{fig:schematic}. In addition to these parameters, uninflated type II particles are also parametrized by the inner annular radius $R_a$.  
The structural and mechanical properties of type I particles are governed by three dimensionless parameters -- $H/t$, $R/t$, and $r/t$ -- which are derived from the system's four geometric parameters.
Type II particles with five defining dimensions are characterized by four dimensionless parameters: the same three as type I particles ($H/t$, $R/t$, and $r/t$) and $R_a/t$.
The experiments select $H = 10~\unit{\milli\meter}$, $R = 10~\unit{\milli\meter}$, $t = 2~\unit{\milli\meter}$, and $0.1~\unit{\milli\meter} \leq r  \leq 4~\unit{\milli\meter}$ for type I particles and $H =10~\unit{\milli\meter}$, $R = 20~\unit{\milli\meter}$, $ 6~\unit{\milli\meter} \leq R_a \leq 12~\unit{\milli\meter}$, $1~\unit{\milli\meter} < t < 4~\unit{\milli\meter}$, and $0.1~\unit{\milli\meter} < r < 3~\unit{\milli\meter}$ for type II particles.  For type I particles, we fix $H/t = 5$, $R/t = 5$, and vary $r/t$.  For type II particles, we fix zero, one, two, or three of the dimensionless parameters at a time, while varying the rest.%\note{we fix different parameters in different experiments. Do we need to specify all the variations?}

The air inlet in type I particles allows airflow into the cavity of the particle, as shown in Fig.~\ref{fig:schematic}A. The central pillar in type II particles has channels connecting the inlet tube to the outer cavity to allow airflow into the particles, as shown in Fig.~\ref{fig:schematic}B. The air inlets in the type I and II particles enable changes in the particle's size and stiffness.  

\subsection{Experimental Measurements of Particle Stiffness}
The stiffness of the particles is measured under uniaxial compression in the direction perpendicular to the cylindrical axis using a materials tester (Instron\texttrademark~$3365$) over a range of inflation pressures $p < 20$~kPa. The particles are inflated by connecting them to a high-pressure air line whose output pressure can be regulated to within $\pm 0.069$~kPa. The inflation at a given $p$ will change the particle radius, which we denote as $R(p)$, as well as all of the other particle dimensions (\textit{i.e.}, $r$, $t$, $H$, and $R_a$). The compressive stiffness $k$ of the particles during inflation is determined from the slope of the force-displacement curve obtained from the materials tester. Specifically, it is calculated based on the compressive force versus the displacement of the compression plate, measured from the point of initial contact with the particle at each fixed inflation pressure. The maximum displacement is $2~\unit{\milli\meter}$, which corresponds to a compressive strain of $\sim 5\%$. We also image the particles during the compression tests at each inflation pressure to determine their size and shape, enabling comparisons to the FEM simulations of particle deformation. 

\subsection{FEM Simulations for Type I Particles}

\begin{figure}[t!]
    \centering
    \includegraphics[width=\linewidth]{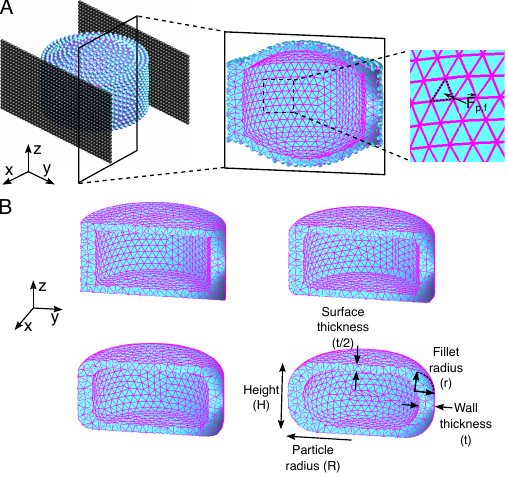}
    \caption{(A) Sketch of the spring network model for (left panel) a type I particle (cyan) with $r/t=4$ compressed between two parallel rough, rigid plates (black) in a direction perpendicular to the cylindrical axis. 
    (center panel) Cross-sectional view of the particle in the left panel, which shows the connections between nodes of the spring network (blue lines) and rough surface of the walls modeled by spheres on a square lattice. (right panel) The inflation pressure is applied through forces ${\vec F}_{p,f}$ split evenly on each node of each triangle $f$ that tessellate the interior surface of the cylindrical shell of the particle. (B) Spring network meshes (blue lines) for type I particles with different ratios of the fillet radius to the wall thickness $r/t$ (from left to right and from top to bottom): $r/t = 0$, $0.5$, $1$, and $2$. The Cartesian coordinate systems in both panels illustrate the particle orientations.}
    \label{fig:springmodel}
\end{figure}

To understand the particle compressive stiffness as a function of inflation pressure, we employ two FEM models to simulate the deformation of type I particles: a hyperelastic model implemented in ABAQUS, and a more general spring network model that allows us to tune the pressure-dependent Young's modulus. We focus on modeling type I particles due to their simple geometry compared to type II particles.

For the first FEM approach, we use an incompressible neo-Hookean model~\cite{treloar_neohooke_1943} in ABAQUS to simulate inflatable particles made from EcoFlex\texttrademark~50. The strain energy density for a neo-Hookean material is
\begin{equation}
    \label{eq:neo-hooke}
    U = C_1 (I_1 - 3),
\end{equation}
where $C_1 = 30$~kPa is proportional to the Young's modulus, $I_1 = \lambda_1^2 + \lambda_2^2 + \lambda_3^2$ is the first invariant of the right Cauchy-Green deformation tensor $\mathcal{F}$, and $\lambda_1$, $\lambda_2$, and $\lambda_3$ are the eigenvalues of $\mathcal{F}$ in descending order. For an incompressible material, $C_1$ is half of the shear modulus in the limit of zero deformation, which yields a Young's modulus of $6C_1 = 180$~kPa at zero strain for EcoFlex\texttrademark~50~\cite{cui_youngsmodulus_2022, pineda_electrofluidic_2015}. In addition, we set the static friction coefficient between the particle and the rigid, rough parallel plates that apply the compressive strain to be $\mu = 0.15$ (Fig.~\ref{fig:springmodel}A). 

Hyperelastic materials can strain soften. Under uniaxial extension, $\lambda_1 = \lambda, \lambda_2 = \lambda_3 = 1 / \sqrt{\lambda}$, where $\lambda$ is the extensional deformation, and Eq.~\ref{eq:neo-hooke} becomes
\begin{equation}
    \label{eq:neo-hooke-uniaxial}
    U = C_1 \left(\lambda^2 + \frac{2}{\lambda} - 3 \right).
\end{equation}
We can then calculate the second derivative of $U$ with respect to $\lambda$, which gives the Young's modulus:
\begin{equation}
    \label{eq:neo-hooke-derivative}
    Y= \frac{d^2U}{d\lambda^2} = C_1 \left( 2 + \frac{4}{\lambda^3} \right).
\end{equation}
For this model, $Y$ decreases with increasing $\lambda$. Note that $Y=6 C_1$ in the limit of zero deformation with $\lambda=1$. 

To vary the degree of strain softening, we developed another FEM-based model for type I particles. We meshed the particles using $N_t \sim 10^4$ Delaunay tetrahedra that include a total of $N \sim 2600$ nodes connected by $N_l \sim 116,000$ linear springs as shown in Fig.~\ref{fig:springmodel}A and B. The potential energy for spring $i$ is
\begin{equation}
    \label{eq:spring}
    U_i = \frac{u_l}{2} \left( a \left(\lambda_i - 1\right)^2 + (1 - a) \left(\sqrt{\lambda_i} - \frac{1}{\sqrt{\lambda_i}} \right)^2 \right),
\end{equation}
where $u_l$ is the characteristic energy scale of the spring, $\lambda_i = l_i / l_{i,0}$ is the stretch of the spring with instantaneous length $l_i$ and rest length $l_{i, 0}$, and $0 \le a \le 1$. The cylindrical shells are meshed such that the average equilibrium rest length is $\langle l_{i,0} \rangle = 1~\unit{\milli\meter}$, which yields a number density of $n_s \sim 10~\unit{\milli\meter}^{-3}$ for the spring networks of type I particles. We set $u_l = 0.018~J$ so that the effective Young's modulus of the spring network for the type I particle is $Y \sim n_s u_l \sim 180$~kPa, which is the same as that for the ABAQUS model. The second derivative of $U_i$ with respect to $\lambda_i$ is
\begin{equation}
    \label{eq:spring-derivative}
   Y_i= \frac{d^2U_i}{d\lambda_i^2} = u_l \left( a + \frac{1 - a}{\lambda_i^3} \right),
\end{equation}
which resembles Eq.~\ref{eq:neo-hooke-derivative}, except that we can tune the local Young's modulus by changing $a$. Here, $a = 1$ corresponds to a linear spring, and $a < 1$ corresponds to a spring that softens upon stretching. 

To simulate purely repulsive, frictional contacts between the inflatable particles and compression plates, we model each rough, rigid parallel plate using a collection of $M$ monodisperse spheres fixed to maintain a square lattice with spacing $d=d_s$, where $d_s$ is the sphere diameter, as shown in Fig.~\ref{fig:springmodel}A. Each node of the spring network within the inflatable particle also includes a sphere with diameter $d_s$ that only interacts with the spheres within the compression plate through the interaction potential: 
\begin{equation}
    \label{eq:wall-interaction}
    U_p = \frac{u_p}{2} \sum_{s=1}^2 \sum_{n = 1}^{N} \sum_{m = 1}^{M} \left(\frac{r^s_{nm}}{d_s} - 1\right)^2 \Theta\left(\frac{r^s_{nm}}{d_s} - 1 \right),
\end{equation}
where $u_p$ the repulsive energy scale, $r^s_{nm}$ is the distance between the $n$th sphere in the inflatable particle and the $m$th sphere in the $s$th wall, and $\Theta(\cdot)$ is the Heaviside step function, which ensures that the node spheres and spheres that make up the walls do not interact when they are not in contact. We set $d_s = 0.75 \langle l_{i,0}\rangle = 0.75~\unit{\milli\meter}$ and $u_p = 5 u_l$. To inflate the particles to pressure $p$, we first identify all $N_f$ triangles in the mesh that cover the interior of the particle. For the $f$th triangle, we assume that the total force resulting from the inflation pressure is
\begin{equation}
    \label{eq:pressure-force}
    \vec{F}_{p,f} = p \vec{A}_{f},
\end{equation}
where $\vec{A}_f$ is a vector normal to the triangle pointing outwards with a magnitude equal to the triangle area. We then evenly distribute $\vec{F}_{p,f}$ on the three nodes associated with the $f$th triangle.

For both FEM-based modeling approaches, we first apply a given inflation pressure, followed by compression tests with the inflation pressure held fixed. In the ABAQUS model, at each $p$, we perform a static analysis with a compression step size of $0.35~\unit{\milli\meter}$ and record the reaction force on one of the compression plates. In the spring network model, we apply the compression quasi-statically. That is, we apply successive compression steps with size $0.35~\unit{\milli\meter}$ to reach a desired total compression. After each compression step, we use the FIRE algorithm~\cite{bitzek_fire_2006} to minimize the total potential energy $U_{tot} = \sum_{i=1}^{N_l} U_i + U_p$ at fixed $p$ until the system reaches a force-balanced state, such that the net force magnitude on each node is less than $10^{-12} u_l / \left\langle l_{i,0} \right\rangle$. In both ABAQUS and the spring network model, we only measure the perpendicular component of the force $\vec{F}$ on one of the compression plates in response to each compression step.  In the spring network model, $\vec{F} = -\sum_{m=1}^M \partial U_p / \partial \vec{r}_m^s$, where $\vec{r}_m^s$ is the position of the $m$th sphere on the compression plate with $s=1$. We determine the particle stiffness $k$ by calculating the slope of a linear fit of $F$ versus the compressive strain up to strains of $10\%$.

\section{Results}
\label{sec:results}

%\begin{figure}
%    \centering
%    \includegraphics[width=\linewidth]{figures/cylinders.pdf}
%    \caption{Stiffness of cylinders as a function of inflation pressure. Cylinders with small fillet radius have a decreasing stiffness with pressure whereas cylinders with large fillet radius show an increasing stiffness with pressure. Each data point is an average of three samples measured three times each. Error bars indicate the standard deviation, dashed line is added to guide the reader.}
%    \label{fig:cylinders}
%\end{figure}

\begin{figure*}
    \centering
    \includegraphics[width=0.8\linewidth]{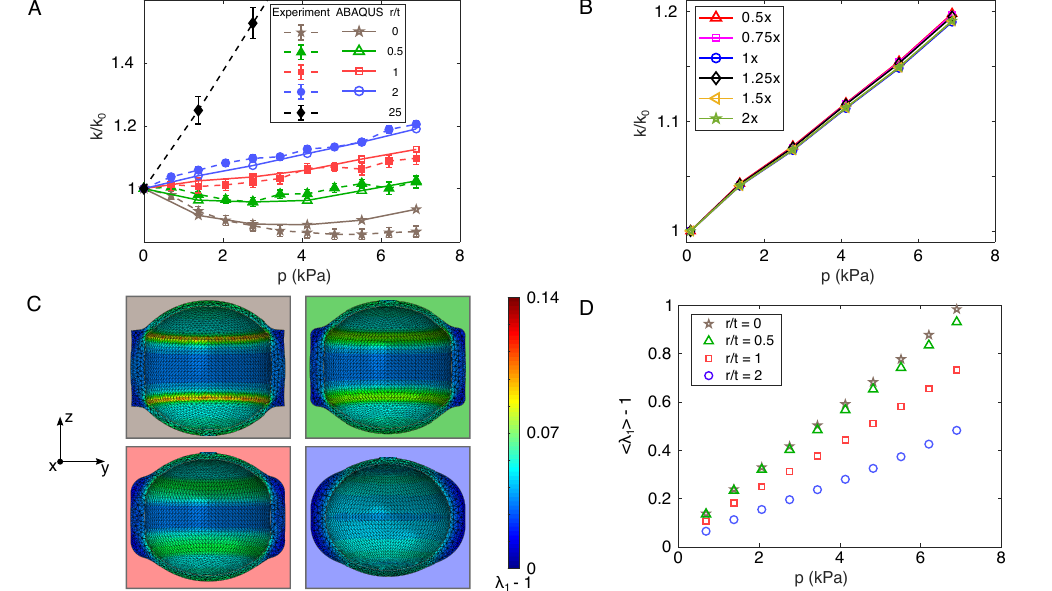}
    \caption{(A) Compressive stiffness $k$ of type I particles with different ratios of the fillet radius to the wall thickness $r/t$ (indicated by color) normalized by $k_0$ at zero inflation pressure plotted versus the inflation pressure $p$ (in kPa). Both the experimental (solid symbols with dashed lines) and ABAQUS simulation results using a neo-Hookean model (open symbols with solid lines) are indicated. Each data point in experiments is an average over three samples measured three times each, with the error bars indicating the standard deviation. (B) $k/k_0$ versus $p$ for type I particles simulated in ABAQUS with the same $r/t = 2$, $R/t = 5$, and $H/t = 5$, but different $t$: $t = 1\unit{\milli\meter}$ (upward triangles), $1.5\unit{\milli\meter}$ (squares), $2\unit{\milli\meter}$ (circles), $2.5\unit{\milli\meter}$ (diamonds), $3\unit{\milli\meter}$ (leftward triangles), and $4\unit{\milli\meter}$ (pentagons). (C) Spatial distribution of the maximum principal strain, $\lambda_1-1$, at $p=6.89$~kPa inflation pressure from the ABAQUS simulations. The particles from left to right and top to bottom have $r/t = 0$, $0.5$, $1$, and $2$. The cross section is through the particle center and perpendicular to both compression plates and the particle's long axis. The Cartesian coordinate system illustrates the orientation of the particles. The color bar indicates values of $\lambda_1 - 1$ for all panels. (D) Mean value of the highest $1\%$ of $\left\langle \lambda_1 \right\rangle - 1$ values in the particle as a function of the inflation pressure for type I particles with different $r/t$ from ABAQUS simulations.}
    \label{fig:abaqus}
\end{figure*}

\begin{figure}
    \centering
    \includegraphics[width=\linewidth]{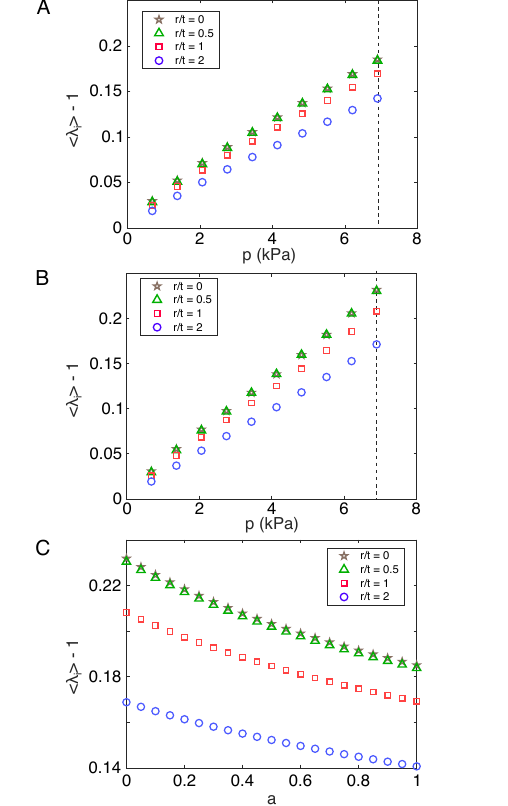}
    \caption{The mean of the highest $1\%$ of strains $\left\langle \lambda_i \right\rangle - 1$ plotted as a function of the inflation pressure $p$ for type I particles with various $r/t$ modeled using spring networks with (A) $a = 1$ and (B) $0$. (C) Mean of the highest $1\%$ of strains $\left\langle \lambda_i \right\rangle - 1$ plotted versus $a$ at $p=6.89$~kPa for type I particles with various $r/t$ modeled using spring networks. The vertical dashed lines in (A) and (B) indicate the pressure used for the data in (C).}
    \label{fig:spring-strain}
\end{figure}

\begin{figure*}
    \centering
    \includegraphics[width=\linewidth]{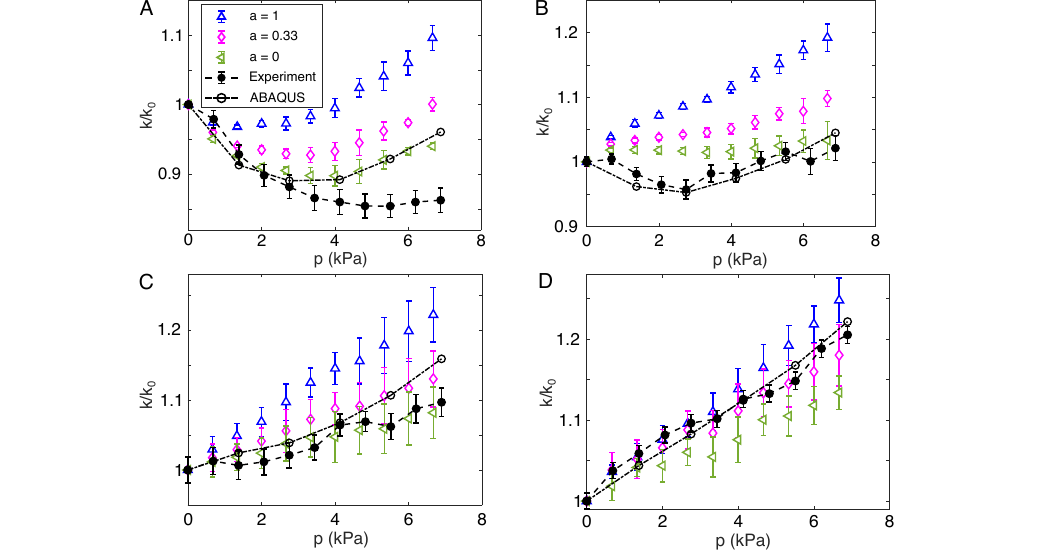}
    \caption{Compressive stiffness $k$ for type I particles plotted versus the inflation pressure $p$, normalized by the stiffness $k_0$ at $p = 0$, at various fillet radius-wall thickness ratios $r/t$: (A) $r/t = 0$, (B) $0.5$, (C) $1$, and (D) $2$. Each panel shows results for the spring network model with $a = 1$ (blue triangles), $0.33$ (magenta diamonds), and $0$ (green left triangles), experimental measurements (black solid circles with dashed lines), and ABAQUS simulations with a neo-Hookean model (black open circles with dash-dotted lines). The 
    error bars are calculated using the standard deviation from $N_r$ independent runs, where $N_r=9$ for the experiments and $6$ for the spring network simulations.}
    \label{fig:spring-stiffness}
\end{figure*}

In this section, we present experimental measurements of the compressive stiffness $k$ for type I and II particles, along with numerical simulation results for $k$ in type I particles, all analyzed as a function of inflation pressure. First, we show that the slope of $k$ versus inflation pressure $p$ can be tuned from positive to negative as we change the internal structure of the type I particles. In particular, the slope of $k$ versus $p$ decreases with decreasing $r/t$, becoming negative for $r/t \lesssim 0.5$ for type I particles. We implement a neo-Hookean model for type I particles in ABAQUS and show that it recapitulates $k$ versus $p$ over a wide range of $r/t$. For particles with negative $k$ versus $p$ slopes, we observe localization of large strains near the corners of the cylindrical particles during compression.  In contrast, when the slope of $k$ versus $p$ is positive, localization of large strains does not occur. Using a more general spring network model, we show that strain softening, arising from the localization of large strains, gives rise to the negative $k$ versus $p$ slopes.  Finally, we find that the dependence of the slope of $k$ versus $p$ on $r/t$ is similar for both type I and type II particles.

\subsection{Compressive Stiffness for Type I Particles}
\label{sec:results-cylinder}

While it was expected that the compressive stiffness $k$ of inflatable particles depends on the inflation pressure $p$, a key finding of this work is that modifying the particle geometry allows us to tune this pressure dependence. In Fig.~\ref{fig:abaqus}A, we show $k(p)$ for type I particles for several values of $r/t$. We find that $k$ increases with pressure for large values of $r/t$, $k$ is nearly independent of $p$ for $r/t\sim 0.5$, and $k$ decreases with $p$ over a large range of $p$ for $r/t=0$. This behavior is in contrast to linearly elastic particles, where $k$ is independent of pressure. 

The ABAQUS simulations of the neo-Hookean model recapitulate the experimental results to within an average normalized root-mean-square error of $\Delta = \sum_{i=1}^{n} (k^E_i-k^A_i)^2/(k^E_i)^2 \sim 0.03$, where $k^E$ and $k^A$ are compressive stiffness values from the experiments and ABAQUS simulations and $n >10$ is the number of measurements. Note that ABAQUS simulations using the Mooney-Rivlin or Yeoh hyperelastic models can better match the experimental results for $k(p)$ at $r/t=0$ when $p > 4$ kPa. We further demonstrate that scaling all of the geometrical parameters by the same factor does not change $k(p)$. Specifically, in Fig.~\ref{fig:abaqus}B, $k(p)$ does not vary significantly when we change $r$, $t$, $H$, and $R$, while keeping the ratios $r/t = 2$, $H/t = 5$, and $R/t = 5$ fixed.

\subsection{Numerical Simulations of the Compressive Stiffness of Type I Particles}
\label{sec:results-sim}

To explain the compressive stiffness $k$ versus pressure $p$ behavior for type I particles for different values of $r/t$ in Fig.~\ref{fig:abaqus}A, we calculated the spatial distribution of the principal strains at each inflation pressure. In Fig.~\ref{fig:abaqus}C, we show how the spatial distribution of the maximum principal strain $\lambda_1-1$ depends on $r/t$.  For large $r/t$, the maximal strain distribution is spatially uniform.  As $r/t$ decreases, $\lambda_1 -1$ becomes non-uniform and the largest maximal strains occur near the fillet. In Fig.~\ref{fig:abaqus}D, we also show that the mean value of the largest $1\%$ of maximal strains in the particle ($\langle \lambda_1\rangle - 1$) increases with decreasing $r/t$. In the previous section, we showed that the most significant strain softening occurs in the $r/t\rightarrow 0$ limit. The results in Fig.~\ref{fig:abaqus}D suggest that the localization of the largest strains near the fillet controls the strain softening (\textit{i.e.}, decreases in $k$ with increasing $p$) in the $r/t \rightarrow 0$ limit.

To further understand the pressure-dependent compressive stiffness, we also carried out numerical studies of the spring network model for type I particles, which allowed us to tune the degree of strain softening during deformation. In Fig.~\ref{fig:spring-strain}A and B, we plot the mean of the largest $1\%$ strain, $\left\langle \lambda_i \right\rangle - 1$, as a function of $p$ for $a = 1$ and $0$, where the parameter $a$ that tunes the degree of strain softening is defined in Eq.~\ref{eq:spring}. We find that $\left\langle \lambda_i \right\rangle - 1$ decreases with increasing $r/t$, for all values of $a$. Therefore, the most substantial strain softening occurs in the $r/t\rightarrow 0$ limit. In addition, we show in Fig.~\ref{fig:spring-strain}C that $\left\langle \lambda_i \right\rangle - 1$ increases with decreasing $a$, which confirms that the most significant strain softening occurs as $a\rightarrow 0$, as well as $r/t \rightarrow 0$.

In Fig.~\ref{fig:spring-stiffness}, we plot $k(p)$  from the spring network model for three values of $a$ and four values of $r/t$ in panels (A)-(D) with comparisons to the experimental and ABAQUS simulation data. We find that the spring network model with $a = 0$ is most similar to the experimental data for $k(p)$ when considering all values of $r/t$ for type I particles. Note that for linear spring networks with $a = 1$, the type I particles exhibit only increasing stiffness with increasing inflation pressure for all $r/t$. These results emphasize that strain softening is necessary to achieve decreasing compressive stiffness $k$ with increasing $p$.  $a = 0.33$ for the spring network model gives the same form for the strain-dependent Young's modulus $Y(\lambda)$ for both the spring network model and neo-Hookean model implemented in ABAQUS, yet the results for $k(p)$ for these two models are slightly different. The deviations are caused by the fact that ABAQUS enforces local incompressibility, whereas the spring network model does not.
 
\subsection{Compressive Stiffness of Type II Particles}

\begin{figure*}[ht!]
    \centering
    \includegraphics[width=\linewidth]{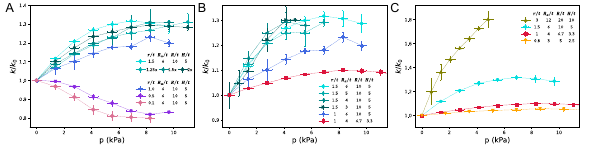}
    \caption{Relative stiffness $k/k_0$ as a function of the inflation pressure $p$ (in kPa) for type II particles with different combinations of $r/t$, $H/t$, $R/t$, and $R_a/t$. In (A), we show $k(p)$ when (circles) all parameters are scaled proportionally at fixed $r/t=1.5$, as well as for (pluses) $r/t=1.0$,  (crosses) $0.5$, and (diamonds) $0.1$ all at fixed $R/t=10$ and $R_a/t=6$. In (B), we show (cyan) $k(p)$ for $R_a/t=6$ (circles), $5$ (diamonds), $4$ (leftward triangles), and $3$ (rightward triangles) at fixed $r/t = 1.5$ and $R_/t=10$. We also show $k(p)$ when two ratios are varied: $R_a/t = 6$ and  $R/t= 10$ (blue pluses) and $R_a/t = 4$ and $R/t = 6.67$ (red squares) both at fixed $r/t=1.0$. In (C), we show $k(p)$ when all four ratios are varied. In all plots, each data point is an average over three samples measured three times each. The error bars indicate the standard deviation and the dotted lines are added as guides to the reader.}
    \label{fig:rt}
\end{figure*}

\label{sec:results-disk}

%The results from the inflation studies of type I particles demonstrate our ability to control their pressure-dependent compressive stiffness $k(p)$. By tuning the geometrical parameter $r/t$, which specifies the initial fillet radius relative to the wall thickness, we can tune $k(p)$ so that it increases with $p$, is independent of $p$, or decreases with $p$. As shown in Fig.~\ref{fig:schematic}, 
Type I particles expand both radially and along the cylinder axis during inflation, whereas type II particles mainly expand radially. Purely radial expansion is desirable for studies of quasi-2D packings of inflatable particles that are easily visualized during applied deformation. In this section, we determine the compressive stiffness $k(p)$ for type II particles. In this case, four ratios, $r/t$, $H/t$, $R/t$, and $R_a/t$, specify the geometry of the particle. We experimentally measure the compressive stiffness versus the inflation pressure for different combinations of the initial values of $r/t$, $H/t$, $R/t$, and $R_a/t$.

%Similar to the discussion of type I particles,
We first demonstrate that scaling all of the geometrical parameters that describe the type II particle shape by the same factor does not change $k(p)$. Specifically, in Fig.~\ref{fig:rt}A, $k(p)$ is similar when we vary $r$, $t$, $H$, $R$, and $R_a$, while keeping the ratios $r/t = 1.5$, $H/t = 5$, $R/t = 10$, and $R_a/t = 6$ unchanged.

Next, we vary one parameter, $r/t$, while keeping the other three parameters unchanged. Similar to the results for type I particles in Fig.~\ref{fig:abaqus}A, the sign of the slope of $k(p)$ is determined by $r/t$ for type II particles as shown in Fig.~\ref{fig:rt}A. At low pressures $(\lesssim 6~\unit{\kilo\pascal})$, $k(p)$ has a strong dependence on $r/t$. For large $r/t~(\geq 1)$, the slope of $k(p)$ is positive; and for small $r/t~(\leq 0.5)$, the slope is negative. At larger pressures $(\gtrsim 7~\unit{\kilo\pascal})$, $k(p)$ reaches a plateau for most type II particles. 
%For type I particles, we only observe a plateau in $k(p)$ at large $p$ for particles with $r/t=0$ that possess strong strain localization. 
We suspect that restricting type II particles to inflate only in the radial direction results in greater radial strains at the top and bottom regions compared to type I particles, leading to increased strain localization.

In Fig.~\ref{fig:rt}B, we keep $r/t$ fixed, while varying the rest of the dimensionless parameters. We find similar behavior for $k(p)$ as that shown in Fig.~\ref{fig:rt}A, \textit{i.e.}, $k$ first increases with $p$ and then reaches a plateau. If we vary $R_a/t$ between $3$ and $6$, while keeping $r/t$, $R/t$, and $H/t$ unchanged, we find no significant difference in $k(p)$. We further observe that while keeping $r/t =1$ fixed, varying the other three ratios alters the behavior of $k(p)$. Although the slope remains positive in both cases, it decreases for smaller values of $R_a/t$, $R/t$, and $H/t$.  

Finally, we examine the case where all four ratios, $r/t$, $H/t$, $R/t$, and $R_a/t$, are varied. As shown in Fig.~\ref{fig:rt}C, $k(p)$ increases with $p$ at small $p$ and the slope of $k(p)$ at small $p$ is mainly determined by $r/t$. However, the four dimensionless ratios that prescribe the initial shape of the particle are important to determine the $p$-dependent compressive stiffness, especially at large $p$. Similar results are found for the evolution of the particle radius $R$ with $p$, as shown in Appendix A.  
%In Fig.~\ref{fig:rt}A and B, we note that the slope of $k(p)$ is positive for cases with $r/t=1$, however, in Fig.~\ref{fig:rt}C, the slope of $k(p)$ for $r/t=1$ is much smaller. 

%\begin{figure}
%    \centering
%    \includegraphics[width=\linewidth]{figures/wall_thickness.pdf}
%    \caption{Side view of two particles with wall thickness $1\unit{\milli\meter}$ (left) and $4\unit{\milli\meter}$ (right). Changing the thickness of top and side walls changes the overall particle shape under inflation.}
%    \label{fig:shape}
%\end{figure}

\section{Conclusions and Future Directions}
\label{sec:conclusions}

In this article, we combined experimental and numerical studies to characterize the properties of two types of inflatable particles with initial cylindrical shapes. We investigated how the initial geometry of the particle affects the pressure dependence of the compressive stiffness $k(p)$. The initial ratio of the fillet radius $r$ to the shell thickness $t$ plays a dominant role in determining $k(p)$. We show that $k$ can decrease, remain the same, or increase with the inflation pressure $p$, for both types of particles, for different values of $r/t$. In particular, $k(p)$ decreases with pressure for small $r/t (\lesssim 0.8)$ and increases with pressure for $r/t \gtrsim 0.8$. We find that at small $r/t$, large strain localization near the ``corners'' of the cylindrical particles leads to strain softening, resulting in a decrease in compressive stiffness with increasing pressure.

This work introduces a novel method for fabricating particles with controllable size and stiffness. 
Our results suggest several promising future directions. First, further investigations of the dependence of $k(p)$ on all dimensionless ratios that characterize the particle shape are necessary to describe the pressure-dependent stiffness, especially at large $p$. Second, it will be interesting to study the structural and mechanical properties of packings of the inflatable particles presented in this work. For example, size and stiffness changes in the particle packings can change the topology of the interparticle contact networks, as well as the force distributions. Tuning the particle size and stiffness of even a single particle can have a significant impact on the contact networks, force networks, and mechanical properties of the packing. One potential application of this study is to use packings of inflatable particles to build quasi-$2D$ robotic granular materials, whose size, shape, and stiffness can change independently and can be optimized to yield packings with specified and adaptable properties.

% Stiffness response of elastomers under strain depends on a variety of interdependent factors including the local geometry, the material moduli, and the stress distributions along different directions in the material. Our results show that we can take advantage of some of this complexity and use it to make inflatable systems whose stiffness behavior can be modified by tweaking certain design parameters. In this work we made cylinders and disk shaped particles which, depending on their geometric features, can either soften or stiffen when inflated. We identified that the ratio two parameters - the curvature of sharp corners (fillets) and the thickness of the walls $(r/t)$ controls the stiffness behavior to the first order. In particular, for large values of $(r/t)$, the stiffness of the system increases with inflation and for small values of $(r/t)$, the stiffness decreases with inflation.

% The focus of this work is inflatable disk shaped particles which can be used to build a quasi-$2D$ robotic granular metamaterial. Using particles that can be inflated or deflated individually, we can create tunable granular packings where local rearrangements can be induced on demand. This directly inspires future work to make shape, size and stiffness changing robotic particles for adaptable granular metamaterials. 
This work raises intriguing questions about the general behavior of stretchable and inflatable systems. It may be useful to conceptualize stress and strain distributions in stretchable sheets through a few key design parameters, which could, in turn, enable inverse design of complex inflatable structures. Balloon-like inflation systems are particularly interesting due to their non-monotonic pressure response to inflation, a phenomenon that has been extensively studied experimentally and theoretically~\cite{stein_inflating_1958, needleman_inflation_1977, muller_inflating_2002, muhaxheri_bifurcations_2024, verron_numerical_2003}. However, the dependence of compressive stiffness on geometric parameters, as demonstrated in this work, represents a distinct effect occurring at smaller strains, separate from previously observed non-monotonic behaviors. Our findings suggest that further research could explore the role of geometry in governing bi-stability and non-monotonic responses in inflatable systems.

Inflatable structures play a central role in soft robotics, serving as actuators. However, current design strategies primarily focus on the final inflated shape rather than the internal stress distribution within the material. Stiffness modulation in soft robotic systems is typically achieved through decoupled mechanisms such as phase change or particle/layer jamming~\cite{buckner_enhanced_2019, hao_low_2022, brown_universal_2010, narang_mechanically_2018, yang_reprogrammable_2021}. Our results suggest a coupled approach, where both shape transformation and stiffness modulation are integrated into a single system through thoughtful geometric design of inflatable components, opening new possibilities for adaptive and multifunctional soft robots. 

% Characterize $k$ versus $p$ for all geometrical parameters. 
% Incorporate size and sitffness changing particles into granular arrays to calculate their mechanical properties using both experiments and FEM modeling. 

\section*{Author Contributions}
N.P.: Methodology, experimental fabrication and measuremnts, data analysis, writing, review and editing, visualization, and project administration. D.W.: Methodology, numerical and finite element simulations, data analysis, writing, review and editing, visualization, and project administration. R.B.: Finite element simulations, review and editing. M.G.: Experimental fabrication, review and editing. M.D.S.: Conceptualization, supervision, and resources. C.S.O.: Conceptualization, supervision, project administration, review and editing, resources, and funding acquisition. R.K-B.: Conceptualization, supervision, project administration, review and editing, resources, and funding acquisition.

\section*{Conflicts of interest}There are no conflicts to declare.

\section*{Data Availability}
Data presented in this article, including data from experiments and FEA simulations are available on GitHub at \href{https://github.com/nidhipashine/Inflatable-particles}{\url{https://github.com/nidhipashine/Inflatable-particles}}

\section*{Acknowledgements}
This material is based upon work supported by the National Science Foundation under the Designing Materials to Revolutionize and Engineer our Future (DMREF) program (Award No. 2118988). This work was also supported by the High Performance Computing facilities operated by Yale’s Center for Research Computing. 
%%%END OF MAIN TEXT%%%

%The \balance command can be used to balance the columns on the final page if desired. It should be placed anywhere within the first column of the last page.

\section*{Appendix A}

\begin{figure}[t!]
    \centering
    \includegraphics[width=\linewidth]{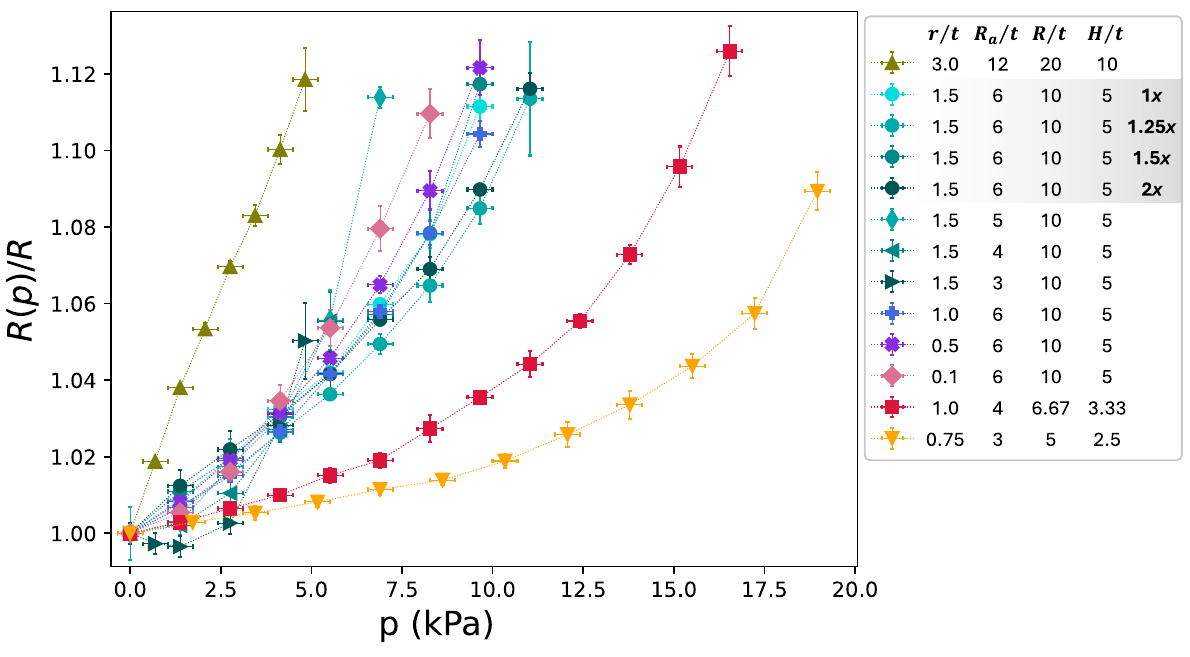}
    \caption{Radius of the particle relative to the uninflated value, $R(p)/R$, as a function of the inflation pressure $p$ for all type II particles studied in Fig.~\ref{fig:rt} with different combinations of $r/t$, $H/t$, $R/t$, and $R_a/t$. Each data point is an average over three samples measured three times each. The error bars indicate the standard deviation and the dotted lines are added as guides to the reader.}
    \label{fig:rp}
\end{figure}

In this Appendix, we describe the relationship between the particle radius and inflation pressure $p$ for all of the type II particles studied in the experiments in Sec.~\ref{sec:results-disk}. Since type II particles mainly expand in the radial direction when inflated, we use the ratio $R(p)/R$ to quantify the change in particle size. As shown in Fig.~\ref{fig:rp}, $R(p)/R$ increases linearly with $p$ for only one type II particle ($r/t = 3$, $R_a/t = 12$, $R/t = 20$, and $H/t = 10$). For the rest of the type II particles, $R(p)/R$ is concave upward, consistent with strain softening. $R(p)/R$ is is most strongly influenced by $R/t$ and $H/t$, however, all four dimensionless shape parameters affect $R(p)$, especially at large $p$.

%\balance

%If notes are included in your references you can change the title from 'References' to 'Notes and references' using the following command:
%\renewcommand\refname{Notes and references}

%%%REFERENCES%%%
 %\bibliography{references, InflatableParticle} 

\end{document}